\documentclass[aps,pra,showpacs,amssymb, nobibnotes,longbibliography]{revtex4-2}

\usepackage[dvips]{color} 
\usepackage {amsmath}
\usepackage[T1]{fontenc}
\usepackage{xcolor,soul}
\usepackage{graphicx}
\usepackage{textcomp}
\usepackage[french]{babel}
\usepackage{graphics}
\usepackage{natbib}
\usepackage[citecolor=red]{hyperref}
\usepackage[normalem]{ulem}

\begin{document}
\title{A linear  Paul trap to study the equilibrium of charged micron-sized  particles in air}

 \author{R. W. Kissangou$^{\diamond}$}
 \author{M.R. Kamsap$^{\dagger}$}\email{mkamsap@yahoo.com}
 \author{M. D. Mboumba$^{\dagger}$}
 \author{M. Houssin$^{\diamond}$}
 \author{C. Champenois$^{\diamond}$}

 \affiliation{$^{\diamond}$Aix Marseille Université, CNRS, PIIM,  Marseille, France}
 \affiliation{$^{\dagger}$Département de Physique, Faculté des Sciences, Université des Sciences et Techniques de Masuku, BP 943 Franceville, Gabon}


\begin{abstract}

  We describe and use a  linear quadrupolar trap to  study the  trapping and equilibrium properties of charged particles. When used to trap atomic ions, these devices   are at the  heart of ion-based quantum computers and optical atomic clocks. To illustrate the characteristics of  trapping a small number of particles and {to} observe   {the various forces acting} on a charged particle  {in equilibrium}, a macroscopic linear trap, operating at room condition,  {traps micron-sized} particles.   {Unlike} atomic or molecular ions, their mass can not be neglected and requires   additional voltage   {to compensate for the force of gravity}.  We   {demonstrate} how this compensation can be  experimentally  {verified} and   {show} that the choice of hollow-core glass spheres   {offers} the advantage of a good  {experimental} reproducibility   {allowing us to determine} the charge-to-mass ratio of the trapped particle with a 13\% uncertainty. In the case of two  particles  {trapped together}, we show how their charge can be deduced from their   {arrangement} along the   {trap's axis of symmetry}.

 \end{abstract}
\keywords{linear Paul Trap~; gravity compensation~; micro-particle~; micro-motion}

\pacs{   }

\maketitle
\section{Introduction}\label{intro}

Paul traps, or quadrupole traps\citep{paul53,paul58,prestage89}, confine ions using time-varying electric fields, enabling a broad range of applications in physics and chemistry, such as precision mass spectrometry\citep{march05}. The development of ion trapping earned Wolfgang Paul and Hans Dehmelt the 1989 Nobel Prize in Physics, and the technology is now ubiquitous. Trapped ions are among the most precise quantum systems for frequency metrology\citep{hausser25,marshall25}, quantum information processing\citep{haffner08}, and quantum simulation\citep{monroe21,fossfeig24}, due to their exceptional control over experimental parameters and the ability to create and manipulate systems with a defined number of individual ions.
When laser-cooled\citep{hansch75,wineland75,neuhauser78}, ions self-organize into ordered structures known as Coulomb crystals\citep{wineland87,diedrich87}. The shape of these crystals depends on the trapping potential’s geometry and the number of ions. In a linear Paul trap\citep{prestage89}, the highly anisotropic potential can cause ions in small clouds to form 1D chains\citep{schiffer93,kamsap17}. In contrast, large ion clouds—containing up to $10^5$ ions in more isotropic potentials\citep{kamsap15a}—form well-defined 3D ellipsoidal systems\citep{turner87,drewsen98,mihalcea23}.

 Macroscopic traps are also used to measure the detailed properties of individual charged particles in the 100-nm to 100-µm size range, including aerosols\citep{yang96,suess99}, liquid droplets\citep{arnold86,kramer99}, solid particles\citep{wuerker59,visan13}, nanoparticles\citep{seo03,cai02}, and even microorganisms\citep{peng04,zhu11}. Additionally, they serve as educational tools for teaching key aspects of charged-particle dynamics and oscillatory motion\citep{winter91,libbrecht18}. Their mechanical analog—based on the equilibrium of a ball on a rotating saddle—also provides a foundation for exploring problems related to equilibrium in rotating frames\citep{rueckner95,fan17}.
Unlike ion traps, macroscopic traps do not require a high-vacuum environment and can operate at atmospheric pressure, where damping is provided by air viscosity. This makes them easy to set up and allows their geometry to be deformed, enabling tuning of their mechanical properties.

In this paper, we demonstrate how a horizontal macroscopic linear quadrupole trap can be used to illustrate the various forces acting on charged particles and to highlight the role of the charge-to-mass ratio in their equilibrium. The paper is organized as follows.
In Section \ref{sec:trap}, we describe the trapping device and explain how air-friction damping facilitates the confinement of macroscopic particles visible to the naked eye. In Section \ref{sec:2part}, we analyze the equilibrium of a two-particle system along the trap’s symmetry axis to illustrate the Coulomb repulsion law and its dependence solely on the particles’ charge. This approach allows us to measure the charge of particles independently of their charge-to-mass ratio, provided they carry the same charge.
In Section \ref{sec:eq_vert}, we examine the equilibrium of a single particle in the vertical plane to explore the balance between gravitational force and the trap’s restoring force, as well as its sensitivity to the $q/m$ ratio. Finally, we show how the trap’s electric center can be identified experimentally.

\section{Trapping charge particles at the atmospheric pressure}\label{sec:trap}
\subsection{Description of the set-up}
\begin{figure}[hbtp]
\centering
\includegraphics[height=5.cm,trim=30 400 30 150,clip]{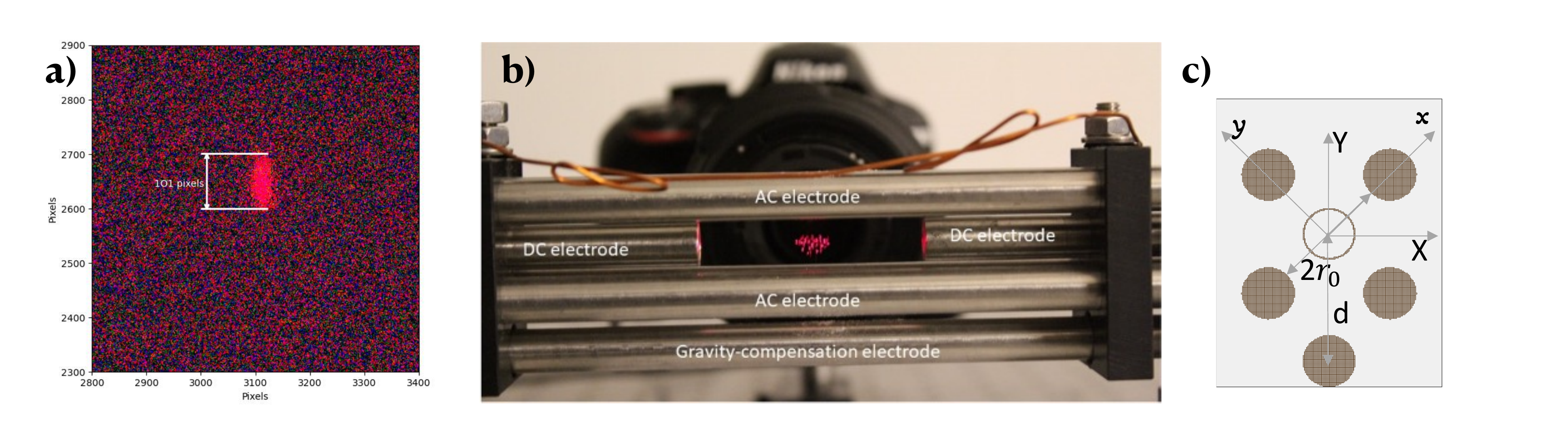}
\caption{The macro-trap with  trapped charged particles : {\bf a):} zoom on the camera screen to focus on one particle trajectory.   {The vertical trace is due to the oscillation of the particle, driven by the oscillating voltage, integrated over the exposure time (see chapter C.)}, {\bf b):} picture of the macro-trap with several charged particles visible by the light they diffuse,   {the distance between the DC electrodes is $2z_0=5$~cm}{\bf c):} geometry of the macro-trap, in the transverse plane.   {The closest distance between an electrode and the trap center $r_0$ is 1~cm and the distance between this center and the gravity compensation electrode $d$ is 2.2~cm}}\label{fig:photo}
\end{figure}

The trapping device used in this work obeys the traditional design of linear quadrupolar radio-frequency traps. It consists of four horizontal cylindrical rods (1~cm diameter) separated by $2r_0=2$~cm connected pair-wise to a time-oscillating voltage at 50~Hz, and two electrodes on the symmetry axis, called "end-cap", separated by $2z_0=$5~cm and  connected to the same DC-voltage, as shown in figure~\ref{fig:photo}.  An additional rod, positioned 2.2~cm below the trap center, can be polarised by an extra DC-voltage $V_g$ to compensate for the weight of the micro-particles. The particles used in the experiments  are engineered hollow glass microspheres sold by 3M under the name "glass bubbles". We chose the K-series for which the particle size (diameter) ranges between 80 and 120 micrometers.  
The particles become charged through the triboelectric effect. This electrostatic phenomenon occurs when two different materials come into contact, resulting in an exchange of electrons at the point of contact between the two materials. To achieve this, we rub the tip of a syringe on a metallic table and then bring it close to the particles to charge them before depositing them in the trap. This technique does not enable the accurate control of particle charge from one experiment to another, but we expect  the triboelectric effect to transfer to the glass beads a charge that is proportional to its surface. We chose  hollow core spheres because the mass is also proportional to the surface, thus reducing the relative dispersion of the charge to mass ratio, compared to the one of the charge and/or the mass themselves. With an uncertainty on the charge $q$ carried by the glass bubbles, operating the trap at an atmospheric pressure of 1 bar is of great advantage. Indeed, the frictional forces caused on the bubbles by collisions with the air molecules ensure a strong damping of the non-driven motion and can extend very broadly the stability zone of the trap \cite{nasse01,vasilyak13}, which allow to easily find efficient trapping. Nevertheless, these particles are light enough to be dragged by any air flow. It is thus advised to build a box around the set-up to be able to observe stable particles during several minutes, turning into hours if the environment is steady.

These particles are observed through the  light they scatter, coming from a red laser diode through the hollow DC-electrodes. The scattered photons are collected by a computer controlled Nikon D3300 camera with a shutter speed range from 1/4000s to 30s. 
During our experiments, the camera is used in a photo mode with an exposure time of 1/30~s.The camera objective is positioned perpendicular to the trap symmetry axis, on the same horizontal plane as the trapped particles as show on figure~\ref{fig:photo}. The camera sensor has 6000 horizontal pixels spanning 23.5 mm and 4000 vertical pixels spanning 15.6 mm, yielding an effective pixel size of 3.9~$\mu $m. Additionally, the detection lens has a measured magnification of $g = 1.33$. The observed displacements in the image can therefore be mapped to the actual displacement of the trapped particle.

To confine the charged particles (charge $q$, mass $m$) in the radial plane, an oscillating voltage of amplitude $V_{ac}$ and frequency $\Omega/2\pi=50$~Hz is applied to the electrodes. They are paired along the diagonals of the square supporting their structure and each pair is polarised with a $\pi$ phase-shifted voltage, compared to the other pair (see fig.~\ref{fig:photo}). 
Our generator delivers two sinusoidal voltages that are exactly opposite to each other, with an amplitude $V_{ac}$ varying from 500 V to 8000 V. To confine along the horizontal symmetry axis of the trap, the end-cap electrodes are polarised with $V_{dc}$ that can vary from 0 to 2000V. The trapped particles at the position $(x,y,z)$ referenced to the geometrical center of the trap,  are then subjected to the following oscillating potential (see Fig.~\ref{fig:photo} for the axis definition): 
\begin{equation}
   \varphi (x,y,t)=\dfrac{V_{ac}\cos (\Omega t)}{r^2_0}\left(x^2-y^2\right) +\dfrac{\kappa V_{dc}}{2z_0^2}\left(2 z^2-x^2-y^2\right) 
\label{eq:phixyzt} 
\end{equation}
where $2 r_0$ and $2 z_ 0$ are the closest distance between facing electrodes in the radial plane and along the symmetry axis, respectively.   The parameter $\kappa \le 1$ is an efficiency parameter which depends on the geometry of the trap.

The large voltages used is another justification for building a box around the set-up. To limit electric hazard, it is important to limit the current that can be provided by the high voltage source. A detailed method to do so can be found in \cite{libbrecht18}. It is very convenient to hold the seven rods with two plastic pieces where the geometry of the trap is fixed, like can be seen on Fig~\ref{fig:photo}~b). One must be careful to keep the set-up long enough so that the trapped particles "see" a small plastic surface. 

\subsection{Damped motion in a linear Paul trap}
The time oscillating electric field allows to circumvent the Earnshaw's theorem stating that charged particles can not be stably confined by a static electric field only. With an appropriate choice of the $V_{ac}$ and $V_{dc}$ voltages, this method of confinement induces two different motions \cite{book_werth}. The obvious motion is the one driven by the time oscillating field, called micro-motion, with an amplitude scaling linearly with the distance to the trap electric center. The fact that the time-averaged kinetic energy of the micro-motion scales with the second order of the distance to the electric trap center is responsible for an effective static potential $\Psi{pp}$, called pseudo-potential, that obeys the same scaling, with $r=\sqrt{x^2+y^2}$ the distance to the trap electric center \cite{dehmelt67} :
\begin{equation}
  \Psi{pp}= \frac{q^2V_{ac}^2}{m\Omega^2}\frac{r^2}{r_0^4}
  \label{eq:pp}
\end{equation}

The motion inside this pseudo-potential, called macro-motion, is strongly affected by the damping effect induced by the background pressure.  It has been shown \citep{nasse01} that this damping can be well represented by a  Stokes law of friction $\vec{F}_S=-6\pi\eta R \vec{v} $ with $R$ the radius of the spherical particle,   $\vec{v}$ its velocity in a medium of viscosity $\eta$.
 It has also been shown in the references \citep{winter91,nasse01} that, for friction coefficient $\gamma=6\pi\eta R$  of the order of $m\Omega$, taking into account viscosity damping  extends the stable region to limit values for $(V_{ac}, V_{dc})$ far greater than in a low pressure environment.  This results in a stable containment of particles with very different mass-to-charge ratios and allows to run the experiments despite the low control on this parameter.   { At atmospheric pressure, the viscosity is large enough to damp the macro-motion to negligible amplitude on a timescale much shorter than the experiment duration.}

\subsection{Micro-motion}\label{sec:micromotion}
\begin{figure}[htb]
\centering
\includegraphics[scale=1]{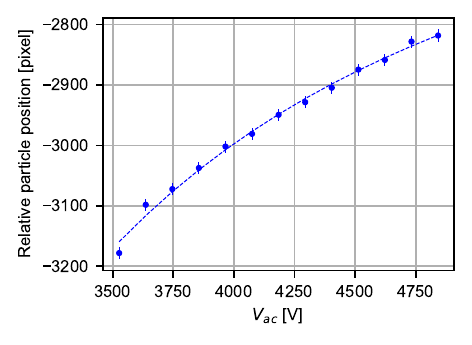}
\includegraphics[scale=1]{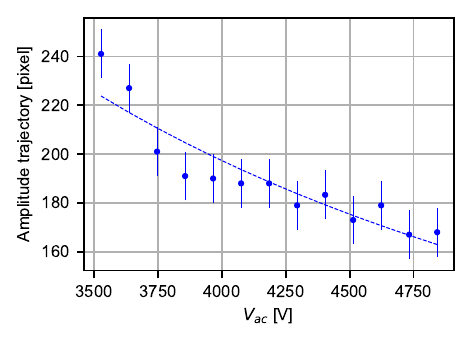}
\caption{  { Trace of a particle trajectory integrated over the exposure time by the camera, for different $V_{ac}$ (1 pixel = 3.9~$\mu$m)}: \textbf{a)}: center of this vertical trace, the dashed line is a $(a/V_{ac}^2+b)$ fit. \textbf{b)}:   {vertical extent on this trace}, the dashed line is a $(a/V_{ac})$ fit.}
\label{fig:micromvt}
\end{figure}

Contrary to atomic ions, the  weight of the charged micro-particles can not be  neglected compared to the confinement force. The equilibrium position of the particles in the vertical direction is then shifted from the electric trap symmetry axis by $Y_0$. This induces a non-negligible driven motion   {(micro-motion)} that is visible on the  picture of Fig.~\ref{fig:photo} {\textbf a)}  as a vertical trace as  the particles oscillate around their equilibrium positions. In a quadrupolar trap, the amplitude of the micro-motion scales linearly with the distance $Y_0$ and with the amplitude $V_{ac}$.
 An analysis of the micro-motion amplitude can therefore be made by measuring the vertical spatial extent of the particle's image when no compensation voltage is applied.
Figure~\ref{fig:micromvt} shows the  trajectory of a trapped particle, integrated over the exposure time,  for different values of  $V_{ac}$. The chosen characteristics are the position of the center of the trajectory, relative to the camera captor (Fig.~\ref{fig:micromvt},\textbf{a)}) and the vertical width of this trajectory (Fig.~\ref{fig:micromvt},\textbf{b)}).

We see  that, as the amplitude $V_{ac}$ increases, the spatial extension of the particle trajectory decreases and in the same time, the center position tends to a higher position on the vertical axis. This can be explained by the competition between the mean  force induced by the pseudo-potential and the gravitation force. Increasing $V_{ac}$ brings the charged particles closer to the trap center and thus, reduces the amplitude of the micro-motion. If we chose $OY$ as the vertical axis oriented positively toward higher altitude with origin on the trap axis, the projection of this force on $OY$ writes
 \begin{equation}
F_Y=- 2 m \left(\frac{q V_{ac}}{ m \Omega r_0^2}\right)^2  Y = - m \omega_y^2 Y
\label{eq_Fy}
\end{equation}
with $\omega_y=\sqrt{2}q V_{ac}/( m \Omega r_0^2)$


Using the force balance on the vertical direction, one can show this behaviour for the displacement $Y_0$ from the trap center:
 \begin{equation}
     Y_0=-\frac{g}{2}\left(\frac{m r_0^2 \Omega }{ q V_{ac}} \right)^2 
     \label{eq:y0}
 \end{equation}
From the fit of the experimental data of figure~2a), we deduce a charge-to-mass ratio  $q/m = (4.3 \pm 1) \times 10^{-4}$~C/kg  and an height of the center of the trap at pixel   {$-2430 \pm 12$} on the camera picture.

The amplitude of the micro-motion  $\cal{E}$ has a different scaling as shown on figure 2b):
 \begin{equation}
   {\cal E} \propto \left(\frac{q}{m} \right)\times Y_0 V_{ac} \propto \frac{mg}{q V_{ac}}
 \end{equation}

\begin{figure}[hbtp]
\centering
\includegraphics[height=6.cm]{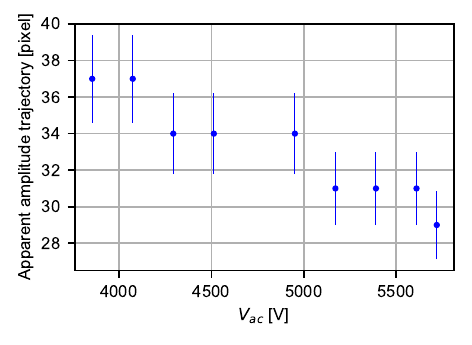}
\caption{ particle image vertical extent   {(1 pixel = 3.9~$\mu $m)} as a function of $V_{ac}$, for a 50~Hz    {illumination}  on the particle. }\label{fig:strombo}
\end{figure}

Micro-motion can be considered as a drawback when one wants to illustrate and analyse different charged particle equilibrium contexts. Cancelling it from the trajectory pictures offers the opportunity to introduce students to the stroboscopic techniques. Stroboscopy is an observation technique that uses periodic flashes of light  to illuminate fast-moving objects at a  frequency which makes the motion appears slower. When the motion is periodic, the perfect matching between the motion frequency and the flash frequency allows to freeze this motion on the pictures.  In our case, we use a gear wheel (a Thorlabs system) with two notches of width 1/20th of a turn, located opposite to each other. The lighting cycle ratio is then 1/10. A suitable motor allows the wheel's rotation speed to be adjusted to illuminate the particles at a rate of 50~Hz (the wheel then rotates at 25~Hz)   
which matches the frequency of the trapping electric field and thus, the particle micro-motion.  We applied stroboscopic illumination to a single trapped particle and analysed how its apparent trajectory extent evolves as a function of $V_{ac}$. The results are shown in the figure \ref{fig:strombo} and a comparison with Fig.~\ref{fig:micromvt}.b)   shows a one order of magnitude reduction. 
In the event that the wheel's rotation is phase-controlled, changing the phase enables one to plot the charged particle trajectory, determine its centre and, consequently, establish the particle average position. Our system doesn't permit the choice of the phase.
In the next section, we focus on the equilibrium properties of charged particles, that can be studied with or without using the stroboscopic techniques. When no stroboscopic tools are available, the center of the trajectory picture can be used for $Y_0$.


\section{Equilibrium between two trapped particles}\label{sec:2part}
In this section, we propose to verify the physical laws that govern the equilibrium, on the horizontal direction, of two charged particles, co-trapped in the same anisotropic harmonic potential. Both particles are subject to the repulsive Coulomb force $\vec{F}_C$ and the axial confinement force $\vec{F}_z$.
The force balance involves only the charge of the particles and is independent from their mass.   {First, we} assume that the two particles carry the same charge $q$. This can be checked by comparing their trajectory center position   {relative to the center of the trap}, within the  uncertainty of the picture analysis. When aligned parallel to the trap symmetry axis, their equilibrium can be described by only one parameter $r_{1,2}$,  the distance between the two particles. The Coulomb repulsion force has an opposite direction on each particle $\{1,2\}$ and its norm is given by the following relation:
\begin{equation}
	|F_{C}| = \frac{1}{4\pi \epsilon_{0}} \frac{q^2}{r_{1,2}^{2}} 
\end{equation}
The restoring trapping force along the trap axis depends on the particle position $z_{1,2}$ relative to the trap electric center, and its norm writes like
\begin{equation}
	|F_{z_{1,2}}|  = \frac{2q\kappa V_{DC}}{z_{0}^{2}}|z_{1,2}|
\label{eq_fz}\end{equation}
 The advantage of a two particle situation is its symmetry that allows the assumption that  $|z_{1,2}|=r_{1,2}/2$. This assumption can be validated by checking the following linear behaviour :
\begin{equation}
	\frac{1}{r_{1,2}^{3}} = \alpha V_{DC}   
\end{equation}
where $\alpha$ depends only on the particle charge, the geometry of the trap and fundamental constants.   {Due to the uncertainty regarding the position of the center of the trap, the two charges may be slightly different within the range of the calculated uncertainty. For two different charges, one can show that $\alpha$ depends on the mean  charge.}

\begin{figure}[hbtp]
\centering
\includegraphics[height=6.cm]{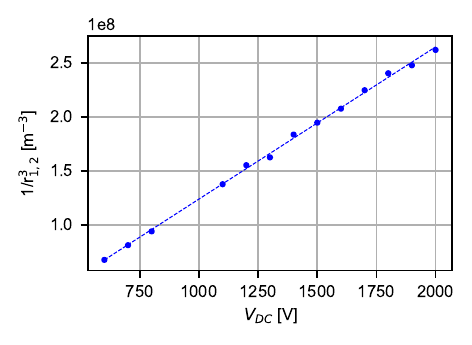}
\caption{ $1/r_{1,2}^{3}$ vs $V_{DC}$ when $r_{1,2}$ is the distance between two co-trapped charged particle.  $V_{DC}$ is the static potential applied to the two end-cap electrodes for confinement along the trap symmetry axis. The dashed blue line is the best linear fit for  the experimental data.}
\label{fig:mesure_d}
\end{figure}

We confine two charged particles aligned with the trap axis and measure their inter-particle distance as a function of $V_{DC}$. The results are presented in Fig.~\ref{fig:mesure_d}, and they confirm that $1/r_{1,2}^{3}$ increases linearly with $V_{DC}$.
They are different ways to go further in the data analysis. First, the comparison of different results based on different particle pairs can give an idea of the variability in the charge carried by the particles. Second, by using a software or a model that compute the profile of the static potential at the bottom of the potential well, it is possible to evaluate the geometric factor $\kappa$ and then deduce the charge carried by each particle. For our trap design, a simulation by the SIMION software \cite{simion} allows to evaluate $\kappa = 0.14$. From the linear slope shown on Fig.~\ref{fig:mesure_d}, we can deduce the charge $q = ((110 \pm 2 )\times 10^{4})e$, 
where $e$ is the elementary charge.
We have conducted this experiment several times and have observed a measurement mean value of about $q = (116  \times 10^{4})e$ and a dispersion of about 7\%. We conclude 
\begin{equation}
 q  = \left((116 \pm 8 )\times 10^{4}\right) e 
 \label{eq:q}
\end{equation}

From equation Eq.(\ref{eq:phixyzt}), we can deduce that the longitudinal oscillation frequency $f_z$ varies from 1.6 to 2.9 Hz when $V_{DC}$ varies from 600 to 2000V. In this experiment, $V_ {ac}= 6450V$ conducting to a   {radial}  oscillating frequency $f_y$ of $17.5$~Hz. The strong anisotropy of the effective trapping potential $ f_y^2/f_z^2 \gg 1 $ justifies the 1D description of the particle equilibrium, parallel to the trap axis.

\section{Equilibrium of a single particle in the vertical direction}\label{sec:eq_vert}
\subsection{Gravitation vs the trapping force}\label{sec:grav}
Analysing the equilibrium of one particle along the trap's vertical axis not only allows to analyse the balance between gravity and a linear restoring force, but it also enables  to determine the charge-to-mass  ratio $q/m$ of a charged particle because the trapping potential is induced by an electric field.

Coming back to the vertical equilibrium condition for a single particle that results in Eq.(\ref{eq:y0}) and considering that the center of the trapping potential well is unknown, we can plot the vertical position relative to the camera captor and use its evolution induced by a change in $V_{ac}$. We then expect
\begin{equation}
Y=-\frac{g}{2}\left(\frac{m }{ q }\right)^2\left(\frac{ r_0^2 \Omega }{  V_{ac}} \right)^2 + \text{cst}. 
 \end{equation}

\begin{figure}[htbp]
\centering
\includegraphics[width=0.5\textwidth]{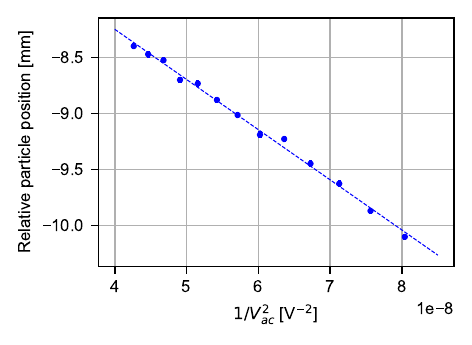}
\caption{Position shift of the center of the trajectory of a single particle as a function of $1/V_{ac}^2$. The blue dashed curve is the best linear fit.}
\label{fig:qsurm}
\end{figure}

The relation between $Y$ and $1/V_{ac}^2$ is thus expected to be linear, as confirmed by the results shown on Fig.~\ref{fig:qsurm}. Knowing all the characteristics of the trap, the value $q/m$ can be extracted from the slope of the curve and we calculate for this data set $q/m=(328 \pm 7)\times 10^{-6}$~C.kg$^{-1}$. From several experiments, we observe   {for $q/m$ a slightly higher mean value of $3.8 \times 10^{-4}$~C.kg$^{-1}$ and a run to run dispersion} of about 13\% conducting to : 
\begin{equation}
    \dfrac{q}{m}=(3.8\pm 0.5)\times 10^{-4}  \  {\rm C.kg}^{-1}
    \label{qsurm}\end{equation}

This charge-to-mass ratio fits in the range mentioned in \cite{thomas15} for lycopodium moss spores and is of the same order of magnitude as the oxide aluminum particles used in \cite{vasilyak13}.

Combining the charge estimated in Eq.~(\ref{eq:q}) and the mean value for $q/m$ given in Eq.~(\ref{qsurm}), we can deduce a typical mass for the glass bubbles :
\begin{equation}
 m = (4.9 \hspace{0.2cm} \pm 0.7)\times 10^{-10} \ {\rm kg}
 \label{eq:m}
\end{equation}

The manufacturer gives a global mean density of 125~mg/cm$^{3}$ for a bottle of micro spheres and a broad range of sphere radius.   { For spheres with a 100$\,\mu$m  diameter, we calculate a mean  mass for a particle} of   {$6.5 \times10^{-11}$~kg which is one order of magnitude smaller} than the calculation deduced from measurements.  {The discrepancy between these two values can be attributed to a incorrect approximation  concerning the packing of the spheres  in the bottle and to  uncertainty regarding the diameter of the spheres.}

\subsection{Adding an extra force to compensate for the gravity}
When the trap is very anisotropic like in III, trapped atomic or molecular ions organise in a chain along the trap axis and this structure is at the core of the quantum computer based on trapped ions \cite{nagerl98,blatt08}. To mimic the equilibrium properties of trapped atomic or molecular ions, it is possible to add an extra vertical force that compensates the gravity. This configuration allows one to observe the self organisation of a few particles  in a chain along the trap axis. 
This extra force is created by an electric field produced by a positive voltage applied to an electrode that is aligned with the centre of the trap but located beneath the quadrupole (see Fig.~\ref{fig:photo}, \textbf{c)}).

\begin{figure}[hbtp]
\centering
\includegraphics[scale=1]{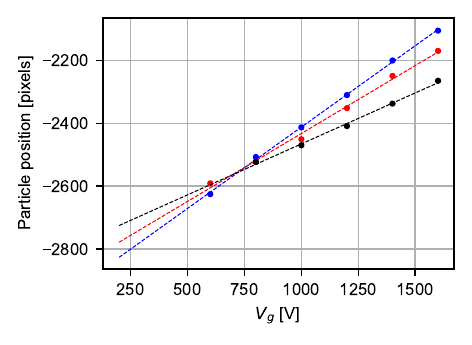}
\caption{Relative particle position   {(1 pixel = 3.9~$\mu $m)} as a function of the potential $V_{g}$ applied to the compensation electrode  for three values of the trapping voltage $V_{ac}$ : 
$V_{ac}=6140$~V (blue dots), $V_{ac}=6790$~V (red dots), $V_{ac}=7940$~V (black dots). The lines are the best linear fit to the experimental data.}
\label{fig:compensation}
\end{figure}

We  present an experimental method for determining the value of the voltage $V_g$ to be applied to the compensation electrode to balance the weight of the trapped particles through an extra static electric field $\vec{E}_c$. The Newton's law to describe the equilibrium of a single trapped particle needs to include a third force $\vec{F}_c=q \vec{E}_c$ proportional to $V_g$.  The restoring force induced by the pseudo-potential trapping is still described by $\vec{F}_Y$ as in Eq.~(\ref{eq_Fy}) :  $\vec{F}_Y=-m\omega_y^2(V_{ac})Y \vec{u}_Y$ with $\omega_y(V_{ac})=\sqrt{2}(q V_{ac})/( m \Omega r_0^2)$  and $\vec{u}_Y$ the unit vector along $OY$ (see Fig.~\ref{fig:photo} \textbf{a)} for the axis definition). The charged particle equilibrium along the $OY$ axis now obeys the following equation : 
 \begin{equation}
  -m\omega_y^2 Y \vec{u}_Y- mg\vec{u}_Y+q\vec{E}_c(V_g)=\vec{0}.   
 \end{equation}
 Assuming that the electric field created by the compensation electrode is uniform in the area explored by the particle,   the new equilibrium position  follows
 \begin{eqnarray}
 Y_0=\frac{q E_c(V_g)-mg}{m \omega_y^2(V_{ac})}.
 \label{eq:position_Vg}
 \end{eqnarray}
 When varying the value for $V_{g}$, for different values of $V_{ac}$, we measure different positions shown on Fig.~\ref{fig:compensation}.
For each value of $V_{ac}$, the linearity of the curve $Y_0=f(V_g)$  confirms the assumption of a local uniformity for $\vec{E}_c(V_g)$. The slopes are different but we observe a crossing of the three curves for the value $V_g^0=730\pm 10$V  and we can conclude that this is the value that cancels $(qE_c(V_g)-mg)$ and set the equilibrium position at the center of the trap. Meanwhile, the electric center of the trap can thus be identified on the camera captor at position -2550 pixels. This position is slightly shifted (300$\mu$m) from the one deduced in \ref{sec:micromotion} from a different set of measurements separated in time by several days.
 
To go further, the slope of the curves in Fig.~\ref{fig:compensation} can be used to determine the $q/m$ ratio. To that purpose, two assumptions guided by the observations are needed :  the  compensation electric field $E_c(V_g)$ can be written as the product of $V_g$ times the nominal electric field $E_{1V}$ created by 1~V applied to the compensation electrode. From Eq.~\ref{eq:position_Vg}, $E_{1V}=mg/(qV_g^0)$. 
 The other  assumption is the dependence of $\omega_y$ with $V_{ac}$ given by Eq.~\ref{eq_Fy}. Combining these two expressions, one can find the equation for the slope : 
 \begin{equation}
     Y_0(V_g)=V_g\times \left(\frac{m}{q}\right)^2\frac{g\Omega^2 r_0^4}{2V_{ac}^2V_g^0}+ cst.
 \end{equation}
The product of each slope with $V_{ac}^2$ is expected to be constant for the same particle, as it is the case for this run of measurements. We find a  5\% discrepancy between the smaller and larger value of this product, that can be attributed to the non-conservation of the charge during the run. Taking this three values into consideration   {including a run to run dispersion}, we can evaluate the ratio
\begin{equation}
 \dfrac{q}{m}=(3.8 \pm 0.4)\times 10^{-4}  {\rm C.kg}^{-1}
    \label{qsurm_2}
\end{equation}
which is compatible with the    {previous} ratio given in Eq.~\ref{qsurm}.
Plugging back this value in the  assumed expression for $E_c(V_g)$, one can get an evaluation for the nominal electric field $E_{1V}$ of $~36$~V/m.

\subsection{Justifying the damping of the macro-motion}\label{sec:damping}

All along this work, we have assumed the damping of the macro-motion by collision with the air molecules. This damping is very well described by a friction Stoke's  force,  like mentioned in \cite{nasse01, libbrecht18}, where the damping coefficient $\gamma$ depends on  the radius of the beads $R$, and  the air viscosity, $\eta$, equal to $1.8 \times 10^{-5}$~kg/(m.s$^{-1}$) at room pressure   :  $\vec{F}_{Stokes} =-\gamma \vec{v}= -6\pi \eta R \vec{v}$. 
  With our measurement of the mass of one particle ($m\simeq 4.9 \times 10^{-10}$~kg) and a supposed mean radius of 50 ~$\mu$m for the beads, we can evaluate the characteristic damping time-scale $(\gamma/m)^{-1}\simeq 1/35$~s. This short time scale confirms why  the macro-motion oscillations are  not observed in our set-up.Concerning the modification of the stability criteria and of the mean pseudo-potential, by the damping effect, this characteristic time $(\gamma/m)^{-1}$ is too large compared to the reduced oscillating field period $\Omega^{-1}$ to have a non negligible impact.

\section{Conclusion}

In this work, we demonstrate how a macroscopic linear trap can be employed to investigate the equilibrium of forces acting on trapped charged particles. The particles used are hollow glass spheres. This shows  the advantage of allowing good reproducibility of the experiments, especially the day-to-day measured charge-to-mass ratio of the particles. The trapped particles have very large masses compared to atomic ions and due to their significant weight, they are trapped below the zero field line, inducing excessive micromotion. This motion is characterised by significant extension of the particle's image, thus making measurements less accurate. We have demonstrated that stroboscopy at the frequency of the driving field can be used to minimise the extensions of the particle's image. We characterised the trapped particle  horizontal and vertical equilibrium in order to experimentally determining its charge-to-mass ratio, but also the charge itself, from which we can deduce a mass, with a 15\% uncertainty. To confine the particles to the trap axis, an additional potential is applied to an electrode below the trap, generating a vertical force on the particles. Including this force in the vertical equilibrium analysis, this method also enables the experimental centre of the trap to be identified.


 \section*{Acknowledgments}
 The authors want to thank J. Pedregosa-Gutierrez for his essential contributions at the early stage of this project. They also thank Guillaume Serin and Audric Husson for their precious help with the set-up. This work has been realized in the frame of the collaboration between Université des Sciences et Techniques de Masuku (USTM) and Aix-Marseille Université (AMU). MRK and RWK would like to thank CNRS for financial support, through the DSCA Africa project, during the completion of this work.

\section*{References}

\end{document}